\documentclass[
letterpaper,
10pt,
final,
balancelastpage,
twocolumn,
superscriptaddress,
floats,
floatfix,
longbibliography,
aps,
pra
]{revtex4-1}
\usepackage{makeidx}
\usepackage{amsmath}
\usepackage{amssymb}
\usepackage{booktabs}
\usepackage{tabularx}
\usepackage{rotating}
\usepackage{acronym}
\usepackage{graphicx}
\usepackage{color}
\usepackage{natbib}
\usepackage[version=3]{mhchem}
\usepackage{tikz} 
\newcommand{\firstprinciples}{\textit{first-principles}}
\newcommand{\abinitio}{\textit{ab initio}}

\newcommand{\denovo}{\textit{de novo}}

\newcommand{\via}{\textit{via}}

\newcommand{\eg}{{e.g.}, }
\newcommand{\ie}{{i.e.}, }
\newcommand{\etal}{\textit{et al.\ }}

\begin{document}

\author{Mojtaba Haghighatlari}
\email{mojtabah@buffalo.edu}
\affiliation{Department of Chemical and Biological Engineering, University at Buffalo, The State University of New York, Buffalo, NY 14260, United States}
\author{Johannes Hachmann}
\email{hachmann@buffalo.edu}
\affiliation{Department of Chemical and Biological Engineering, University at Buffalo, The State University of New York, Buffalo, NY 14260, United States}
\affiliation{Computational and Data-Enabled Science and Engineering Graduate Program, University at Buffalo, The State University of New York, Buffalo, NY 14260, United States}
\affiliation{New York State Center of Excellence in Materials Informatics, Buffalo, NY 14203, United States}

\title{Advances of Machine Learning in Molecular Modeling and Simulation}

\begin{abstract}
In this review, we highlight recent developments in the application of machine learning for molecular modeling and simulation. After giving a brief overview of the foundations, components, and workflow of a typical supervised learning approach for chemical problems, we showcase areas and state-of-the-art examples of their deployment.
In this context, we discuss how machine learning relates to, supports, and augments more traditional physics-based approaches in computational research. We conclude by outlining challenges and future research directions that need to be addressed in order to make machine learning a mainstream chemical engineering tool.

{\centering
\includegraphics[width=0.4\textwidth]{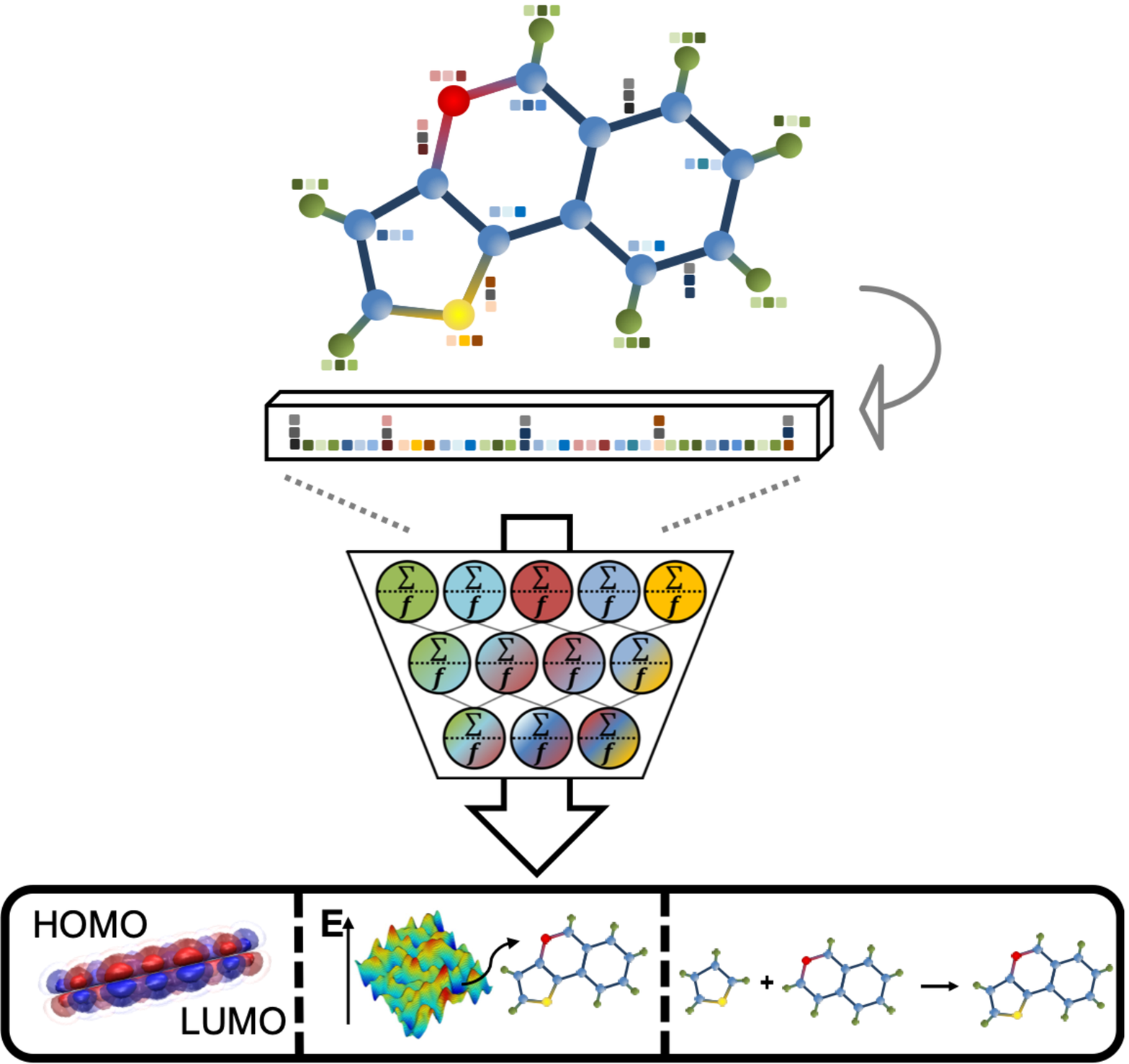}
\par
}

\end{abstract}



\maketitle


\section{Machine Learning from a Chemical Perspective}
\label{sec:introduction}
Over the past few years, data science has started to offer a fresh perspective on tackling complex chemical questions, such as discovering and designing chemical systems with tailored property profiles, revealing intricate structure-property relationships (SPRs), and exploring the vastness of chemical space \cite{nsfreport}. Data-derived prediction models serve as surrogates for physics-based models that are at the heart of traditional modeling and simulation work. They are attractive, because they are usually dramatically less demanding than physics-based models and can thus be deployed in studies of correspondingly larger scope and scale. If trained on experimental data, they are also not subject to the approximations made in physics-based models and may thus not exhibit the resulting discrepancies with respect to non-idealized experimental findings. Of course, data-derived models have their own intrinsic errors and limitations, which we will address in the course of this review.

Machine learning (ML) is a data mining technique and used to create data-derived models. It enables us to extract complex and often hidden correlations (and thus ideally insights, patterns, rules, and guidance) from given data sets and to encapsulate them in mathematical form. ML is commonly categorized into four types, \ie\ supervised, semi-supervised, unsupervised, and reinforcement learning. The main difference between these types is in essence the amount of information (\ie\ labeling, context) that is available for the target variable that serves as the ground truth for the training of an ML algorithm. While all ML types have found application in chemical research \cite{Alberi2018}, supervised learning has so far been most commonly used, and this review will thus focus on it.
The popularity of supervised learning may be due to its heuristic and intuitive approach to learning, which is similar to a scientist's way of gaining insights into SPR. A supervised prediction model can be thought of as a function $f \colon X \rightarrow Y$ that maps an input $x \in X$ to an output $y \in Y$, where $x$ in this context is the feature representation of a chemical system and $y$ its target property. If the variables $x$ and $y$ are continuous (numerical), then the mapping is a regression; if they are discrete (categorical), then it is a classification. 

We can utilize a host of supervised ML algorithms to train and optimize model $f$ to approximate the output value for a given input. 
Two popular algorithms that have been widely used are artificial neural networks (ANNs) and kernel methods. Both can be thought of as transforming the input $x$ into a new feature (latent variable) space, in which it becomes linearly correlated with the output $y$ \cite{Rupp2018}. The transformation itself is typically highly non-linear. A major advantage of the ANNs is their capacity to transform features sequentially through several layers, which is referred to as deep learning. Kernel methods, on the other hand, usually transform features in a one-step process using kernel functions. Unlike in ANNs, this process is predefined prior to the tuning of the model's parameters, and is thus less flexible to learn the best latent variable space. The advantage of kernel methods is their superior performance in finding global solutions, even for small-size data sets where ANNs have deficits.
The support vector machines and kernel ridge regression are two common examples of kernel-based algorithms.

The relationship between a molecular structure and its properties is deterministic, \ie\ there exists an exact mapping from fundamental physics (\ie\ the Schr\"{o}dinger equation). This mapping is ultimately the foundation for traditional modeling and simulation techniques. The topology of ML models is generally very flexible (as, \eg\ expressed in the universal approximation theorem for ANNs), so that they can learn and recover the underlying SPRs of a problem, even from simple chemical representations (assuming no significant loss of information within this representation).

We can consider a feature representation method as a function $g : M \rightarrow X$ that maps a basic chemical representation $m \in M$ to a feature input $x \in X$ (typically called a descriptor). 
The representation $m$ may contain spatial or at least topological information that defines a molecule and is expressed, \eg\ in atomic coordinates, simplified molecular-input line-entry system (SMILES) \cite{Weininger1988}, international chemical identifier (InChI), or other formats. 

A common feature space $X$ is spanned by structural descriptors.
Some ML approaches also utilize physical or (physico-)chemical properties as descriptors, such that $g$ corresponds to a simulation or some other type of calculation (including those from \textit{first principles}). As this approach builds physics into the feature space, it has a certain appeal and has gained corresponding popularity. However, it is important to point out that the computational cost of obtaining such descriptors (which include optimized geometries) may easily make this the bottleneck of an ML approach, in which case it will limit its utility as an efficient surrogate for the prediction of $y$. This issue has to be considered as part of a cost-benefit analysis.

Another class of descriptors is designed to capture the local environment of each atom in a molecule \cite{Bartok2013c}. This approach considers a molecule as a graph with atom and bond (\ie\ node and edge) features. Each atom can interact with all other atoms in its immediate vicinity, which results in an update of the corresponding local atomic features. Incidentally, this approach has its roots in both chemical and data sciences: In the context of molecular simulations, cutoff radii have long been used to exploit the short-ranged nature of intermolecular interactions. In data science, the idea of dynamic irregular graphs provides the underpinning of graph convolutional neural networks. The overlap of the two disciplines in this area has led to many methodological advances for the generation of descriptors. Results from a number of recent studies suggest that an ensemble of local features (rather than a global representation), is able to provide a more robust solution to the challenges involved with variant graph size and the order of atoms in molecules \cite{Kearnes2016b, Hy2018}.

The descriptors discussed so far are essentially hand-crafted 
to explicitly expose certain structural, physical, or (physico-)chemical information $x$ from $m$ and provide a structured (\ie\ tabular) feature representation. 
Alternatively, the feature generation $g$ can also be merged into the prediction model $f$ and both will be jointly optimized, \eg\ through hidden layers (latent space) in deep learning. This class of descriptors is called learned features \cite{Schutt2016a}. 

\begin{figure}[tbp]
    \centering    \includegraphics[width=0.48\textwidth]{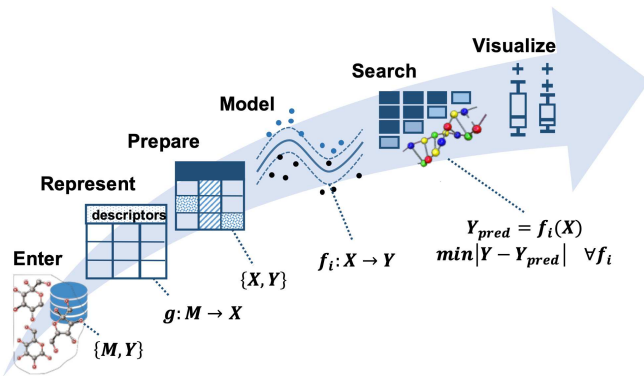}
    \caption{The major tasks and mathematical setup of a supervised machine learning workflow: For a given data set $\{ M , Y \}$, in which for a number of molecules in basic chemical representation $m \in M$ the target property $y \in Y$ is given, we apply a feature representation method as a function $g : M \rightarrow X$ that maps $M$ to a feature input space $X$. After cleaning and other preprocessing steps, we use $\{ X , Y \}$ to formulate an ML model $f \colon X \rightarrow Y$ that maps the feature input space $X$ to the output label space $Y$. The ML model is trained on the training subset of $\{ X , Y \}$, and subsequently validated and optimized on its testing subset, so that it minimizes the prediction error for $Y$.}
    \label{fig:workflow}
\end{figure}

The overall ML workflow for chemical problems encompasses a number of steps as shown in Fig.\ \ref{fig:workflow}, including parsing, cleaning, and preprocessing a chemical data set $\{ M , Y \}$, compiling an array of descriptors \via\ $g$, as well as training, evaluation, and validation of the prediction model $f$.

\section{Applications of Machine Learning in Chemical Research}
\label{sec:data}
In the following section, we summarize three application areas of ML in chemical research, with particular consideration of the inherent structure of the associated data sets, types of representation, and connections to traditional modeling. 
We limit the scope of our discussion to molecular systems, which still cover a broad range of use cases.

\subsection{Discovery and Design of New Compounds}
\label{subsec:compounds}
The application of ML for the exploration of chemical space and the creation of new compounds (ranging from small molecules to polymers and materials) can be divided into two distinct approaches: (i) \textit{discovery}, \ie\ ML is used to generate fast prediction models for properties of interest, with which large-scale surveys of chemical space can be conducted in order to identify compounds that exhibit desired property profiles; (ii) \textit{design}, \ie\ ML is used to develop a quantitative understanding of the SPRs of interest, which can be inverted to pursue the targeted, rational design (or inverse engineering) of compounds with particular properties. While the core activity, \ie\ the ML of SPRs, is the same in both cases, its use follows different perspectives. 

\begin{figure}[tbp]
    \centering    
    \includegraphics[width=0.48\textwidth]{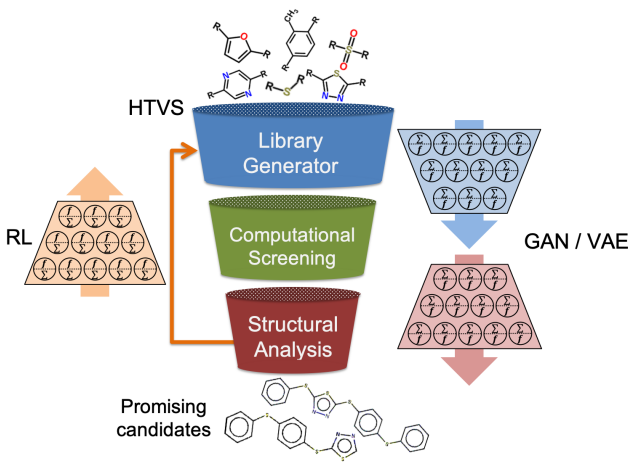}
    \caption{Flowchart showing a computational funnel typical for high-throughput virtual screening (HTVS) studies. The neural network schemes on the left and right represent deep generative model architectures that can conceptually replace different elements of the screening funnel. Both generative adversarial networks (GANs) and variational autoencoders (VAEs) include two networks that revolutionize the conventional generation and analysis steps by probabilistic means. A deep reinforcement learning (RL) network can also be trained to bias the generation towards promising candidates.}
    \label{fig:htps}
\end{figure}

\textit{\textbf{Discovery.}} The idea of employing data-derived prediction models instead of physics-based models (or experimentation) as a means to characterize candidates in the search for new molecules 
may be one of the earliest applications in chemistry, for which the use of ML was proposed. Traditional molecular modeling and simulations have been used for this purpose for many years. More recently, they have also been employed in the context of virtual high-throughput screening studies, in which they are tasked with assessing entire libraries of candidate compounds (see Fig.\ \ref{fig:htps} and, \eg\ Ref.\ \cite{Hachmann2011,Hachmann2014}). However, the computational footprint, in particular of  \firstprinciples\ approaches, is limiting both individual as well as large-scale studies that seek to identify compounds with specifically targeted properties.

The application of data-derived prediction models enables us to dramatically accelerate the survey of chemical space, often by several orders of magnitude. A speed-up of that magnitude allows a corresponding increase in the scale and scope that is viable for screening efforts. (It is thus sometimes referred to as \textit{hyperscreening}.) The candidate libraries are typically generated from a collection of moieties and patterns that are of interest in a given context \cite{Gong2018}. The combination of such a set of building blocks leads to a molecular library for a particular domain in chemical space, \ie\ the candidates belong to the same distribution \cite{pyzer-knapp2015}. A number of experimental or high-level computational training sets have been developed for specific classes of molecules \cite{Butler2018, Li2018b}. Since these data sets focus on relatively similar compounds from the same distribution, the choice of representation and the ML model training are arguably less challenging compared to more universally applicable models. 
The extrapolative use of data-derived prediction models outside the domain for which they were trained has to be conducted with great care and caution, as they are least reliable here. This is a conceptual challenge, as screening studies are often interested in compounds with extreme properties that are likely at the margins of the distribution, where the predictions are least reliable. Iterative retraining of ML models allows us to shift the training data distribution into particular areas of interest, thus making them more robust for use in discovery.

A reasonably diverse collection of molecules can be found in the open-source QM9 data set originally extracted from the GDB-17 chemical universe of 166 billion organic molecules \cite{Ramakrishnan2014}. The QM9 data comprises computed geometries and properties for 134,000 molecules at density functional theory (DFT) level. Due to the diversity of molecular structures and broad range of calculated properties, QM9 plays an important role as a benchmark data set for new models and methods \cite{Ferre2017, Collins2018}. Its contribution to method developments can be compared to the MNIST data set in the hand-written character recognition community \cite{LeCun2013}. 
In contrast to, \eg\ data sets from \firstprinciples\ modeling, those from data-derived models have so far rarely been used for the generation of new reference data. Yet, they have played an important role in a number of methodological advances in the field. As a result, the reported accuracies for many of the recent ML prediction models surpass those of traditional molecular modeling and simulations \cite{Faber2017a}.

\textit{\textbf{Design.}} While discovery is still based on a traditional trial-and-error process -- albeit one drastically accelerated 
by ML -- the notion of a deliberate, \denovo\ design of new compounds represents a different research paradigm.
It addresses the problem that even rapid and efficient hyperscreening studies can only scratch the surface of the practically infinite molecular space.
Instead, the design paradigm seeks to utilize insights into the SPRs obtained from ML for the targeted creation of systems with specific properties.  
The understanding of how changes in the molecular structure (or a compound's features) lead to changes in the desired properties can be inverted to gain a property to structure mapping. 
The mathematical structure of a data-derived SPR prediction model (\eg\ the dominant features, principal components, latent variables, or learned features) yields a foundation for inverse design.  
Models that are less easy to interpret can be projected onto surrogate models for which the extraction of guidelines is easier.
A key challenge is to realize the simultaneous enhancement of different properties. 
The emerging design rules can be used to formulate individual compounds \cite{Sanchez-Lengeling2018}, but also to identify high-value domains in chemical space, which can be enumerated in screening libraries (\eg\ by sampling compounds similar to a lead compound). The latter approach is effectively interfacing the discovery and design perspectives and allows both physics-based and data-derived modeling studies to be more targeted.

Another approach that is very promising and has caused much excitement is the application of generative models (see Fig.\ \ref{fig:htps}). 
For instance, Sanchez-Lengeling \etal\ have shown that a generative adversarial network (GAN) in tandem with reinforcement learning can outperform evolutionary algorithms in order to bias the generative process toward the extreme regions of a property distribution \cite{Sanchez-Lengeling2017, Putin2018}. 
The use of GANs for molecular design and library generation is very recent and a number of concerns and challenges still need to be overcome in their development. 
Two of the principal challenges of GANs (and other generative approaches) are the rate at which invalid (\ie\ chemically irrelevant or non-sensical) structures are generated,
and their ability to produce topologically different molecules compared to the underlying training data \cite{Kadurin2016, Popova2018}.  
Another example of generative models are variational autoencoders (VAEs) that learn the distribution of embedded space and thus enables tuning in that space \cite{Gomez-Bombarelli2018}. 
Recurrent neural networks (RNNs) operate in a sequential manner similar to creating new molecules one atom at a time. One benefit of RNNs is their memory mechanism that allows them to remember the effects of previous sequences \cite{Segler2018}.   

\subsection{Creation of New Modeling Techniques}
\label{subsec:mlff}
Instead of \textit{replacing} physics-based with data-derived modeling entirely as outlined in Sec.\ \ref{subsec:compounds}, ML can also be used to (i) \textit{calibrate} and \textit{correct} the results of physics-based models to account for some of their systematic errors; (ii) \textit{complement} traditional modeling and simulation approaches (\ie\ employ combinations of physics-based and data-derived models); and (iii) facilitate the \textit{development of new} physics-based modeling techniques.  

The \textit{calibration} approach allows us to improve the predictive performance of physics-based models and obtain high-quality results at the cost of lower-quality methods. It can also help bridge the gap between experiment and theory that results from the inherent approximations in the latter (see, \eg\ Ref.\ \cite{Hachmann2014}). Transfer learning is an ML design methodology that has been a particularly successful technique in this context \cite{Goh2017a, Sultan2018a}.
In the \textit{combination} approach, we only utilize ML for aspects for which no good physics-based models are available or where their use is impractical (\eg\ because of insufficient accuracy, prohibitive cost, or other numerical issues). We thus retain as much of the physical foundations and robustness of traditional modeling as possible, while being pragmatic about the parts of a problem, where that is not possible (see, \eg\ Refs.\ \cite{Afzal2018, Ribeiro2018, Wei2018}). 

The \textit{development of entirely new modeling techniques} by means of ML has seen encouraging pioneering efforts, in particular for force fields (FFs) and DFT.
The major driving force behind ML-generated FFs is the lack of generalizability in the classical FFs and the interatomic potentials that underpin them. 
This is an area where ML is apparently able to bridge the accuracy and versatility typically seen from quantum chemistry and the efficiency of molecular mechanics simulations. 
A recent line of research has focused on learning interatomic potentials from quantum chemical data sets \cite{Khorshidi2016}. 
There are two specific challenges involved in this application that make it distinct from prediction models for molecular properties. 
One is the need for a diverse sampling of non-equilibrium chemical conformations, as
both ML and classical FFs perform poorly outside of their applicability domain. 
Access to a diverse collection of high-quality training samples is thus essential in creating ML FFs. 
For instance, Botu \etal\ have improved on previous work by diversifying their training data, \eg\ by adding more atomic environments and applying clustering methods \cite{Botu2017}. 
Smith \etal\ have pushed the normal mode sampling method to obtain single point energies for more than 20 million conformations generated for 58,000 small molecules \cite{Smith2017b}. 
The results of these efforts were shown to be efficiently generalizable, even for the simulation of more complex phenomena. 
The second important challenge is to conserve the consistency between potential energies and forces as discussed by Chmiela \etal\ \cite{Chmiela2017a}. 
They provide a robust solution to this challenge by developing gradient-domain ML models (which reproduce global FFs by training in the force domain and incorporating both energies and forces) in an automated fashion, thus learning accurate ML FFs. 

In the DFT context, ML is used to create new functionals for different terms in the electronic Hamiltonian.
The exact form of several functionals (\eg\ the kinetic energy functional for interacting electrons or the exchange-correlation functional in the Kohn--Sham formalism) are unknown and otherwise approximated by physical reasoning. 
The ML-generated functionals allow DFT to avoid common failures, such as in accurately describing bond-breaking processes. Different ML functionals for specific classes of molecules, target properties, and electronic structure situations are being developed, as are fast methods that, \eg\ learn energy functionals directly without having to solve the Kohn--Sham equations, thus making them a viable approach for \abinitio\ molecular dynamics simulations \cite{Brockherde2017}.

\subsection{Predictions of Chemical Reactions and Catalyst Systems}
\label{subsec:catalysis}
Research on chemical reactions is another field that has been benefiting from the advances in ML methodology. 
ML has been paving the way for a better understanding of chemical transformations with numerous real-world implications. SMILES are often the representation of choice for both the inputs (reactants and reagents) and outputs (products) of data-derived models for chemical reactions.
These models are trained on known reactions to recognize structural patterns that may undergo bond-breaking or -formation in the course of a reaction or catalytic process \cite{Coley2018b}. 
One particularly important data set for this application domain is the result of the US patent reaction extraction by Lowe \cite{Lowe2014a}. 

The progress in predicting organic reactions and their products has been particularly noteworthy in recent years.
Nam \etal\ have introduced sequence-to-sequence models to address the reaction prediction task similar to linguistic translation problems \cite{Nam2016}. 
More recently, Schwaller \etal\ outperformed a similar approach in an end-to-end template-free model with a focus on the attention mechanism and a new tokenization strategy \cite{Schwaller2018}. 
Coley \etal\ introduced a graph convolutional neural network approach with competitive performance. It was used for the  prediction of reaction products as well as reactive sites of the reagents that are most likely to initiate a reaction \cite{Coley2019a}. 
One major contribution of the last two studies is the development of web applications to facilitate easy access to their models. These tools are available \via\ the IBMRXN and ASKCOS websites, respectively \cite{ibmrxn2018, askcos2018}. 

A promising direction of ongoing work is the prediction of reaction pathways and mechanisms. 
All these efforts ultimately aim for a practical and more generalizable implementation of retrosynthetic analysis, which has been a grand challenge in organic chemistry for many years \cite{Klucznik2018}. Insights regarding the synthetic feasibility of virtual compounds are also a key concern for the screening library generation and molecular design efforts discussed in Sec.\ \ref{subsec:compounds}.

\section{Outlook on Future Directions}
\label{sec:future}

\subsection{Feature Representations}
\label{subsec:feat}
As discussed in Sec.\ \ref{sec:introduction}, the descriptors of a given molecular system are an abstraction of its detailed nature (as well as a numerical representation). 
The choice of a suitable feature space is still our first and most effective means to infuse physics into ML models.
There have been efforts to define criteria for the development of efficient descriptors \cite{Bartok2013c}, \eg\ that they are (1) invariant to the symmetries of the underlying physics; (2) easy to interpret; (3) expressed in a direct and concise form to avoid redundancy and the curse of dimensionality; and (4) computationally efficient.
However, developing molecular representations that adhere to all these criteria has been an exceedingly difficult task.
More importantly, there is now agreement that ML approaches may require different types of descriptors to recover the entirety of SPRs of molecular systems.  
Further research into the creation of new descriptors (including fingerprint schemes) as well as the formulation of additional criteria will be necessary for the foreseeable future.
The accessibility and flexibility of deep learning models can accelerate future developments \via\ learned features and theory-informed models.

\subsection{Machine Learning for Small Data}
\label{subsec:design}
While ML ideas became popular during the recent 'big data' wave (\ie\ in chemistry with the emergence of large-scale screening result from high-level \firstprinciples\ modeling), large data sets are in practice more often than not unavailable.
In fact, problems for which data is (still) sparse tend to be of particular interest.
As the data generation (both from experiment and modeling) is often a limiting factor, we will have to strive to reduce its cost or the number of data points needed to obtain ML models of a desired accuracy.  
It is thus essential to put an emphasis on developing ML methods that achieve better performance on small data sets. As mentioned in Sec.\ \ref{subsec:mlff}, transfer learning is a promising approach in this context.
We will also need to employ smart sampling methods and identify data points that are most important for the training of ML models.
Active learning strategies offer a path towards this goal \cite{Hase2018a, Tran2018, Gubaev2019}. 
Many of these techniques are of general-purpose utility, but some will have to be tailored towards the specific problem settings of data-derived models for chemistry.

\subsection{Software and Tool Development}
\label{subsec:software}
The idea to utilize ML and other data mining techniques in the chemical domain is so recent that much of the basic infrastructure has not yet been developed, or is still in its early stages \cite{nsfreport}. 
The majority of tools and expertise tend to be technically involved, labor intensive, or otherwise unavailable to the community at large. 
Many researchers are now starting to pursue open-source software development projects to tackle this situation \cite{Hachmann2018}. 
However, the lack of rigorous development guidelines remains a challenge that researchers from domain science need to overcome to make their efforts lasting and sustainable. 
The Molecular Sciences Software Institute (MolSSI) is one of the pioneers in establishing best practices and guidance for early-stage software developments in this field \cite{Krylov2018, Wilkins-Diehr2018}.

\section{Conclusions}
\label{sec:conclusions}
In this review, we discussed how ML can advance traditional modeling and simulation by (partially) replacing them (\ie\ choosing data-derived over physics-based models or combining the two); calibrating, augmenting, or otherwise correcting their results; targeting studies and their objectives; and providing the means to effectively mine their results for a deeper understanding of hidden SPRs. Many ML models are still built on data provided by modeling and simulation -- often as part of virtual high-throughput screening studies --
and combining ML and traditional modeling infuses physics and robustness into the resulting data-derived prediction models. These and other emerging ML techniques have been enabling accelerated discovery and rational design in numerous areas of chemistry. Its early successes indicate that ML is bound to become a mainstream tool in chemical research. Yet, there is still much to (machine) learn on how to develop the full potential of ML in chemistry. 

\section*{Competing Financial Interests}
The authors declare to have no competing financial interests.

\begin{acknowledgments} 
MH gratefully acknowledges support by Phase-I and Phase-II Software Fellowships (grant No.\ ACI-1547580-479590) of the National Science Foundation (NSF) Molecular Sciences Software Institute (grant No.\ ACI-1547580) at Virginia Tech.
JH acknowledges supported by the NSF CAREER program under grant No.\ OAC-1751161, the NSF Big Data Spokes program under grant No.\ IIS-1761990, and funding by the New York State Center of Excellence in Materials Informatics (grant No.\ CMI-1148092-8-75163).
\end{acknowledgments}

\section*{Annotations}

    \begin{itemize}
      \item \cite{nsfreport} This NSF workshop report compiles the opinions of a group of active researchers in the field regarding the current challenges and future opportunities offered by data-driven approaches in the chemical domain.
    
      \item \cite{Sanchez-Lengeling2018} This review discusses the recent advances in inverse molecular design using deep generative models.
      
      \item \cite{Smith2017b} In this study, deep learning is used to fit interatomic potentials and develop the so-called ANI model for transferable data-derived potentials with comparable accuracy to the reference DFT calculations.

      \item \cite{Coley2018b} This study surveys the role of ML in synthesis planning and the prediction of reaction outcomes.
      
      \item \cite{Hachmann2018} This paper presents a software ecosystem for the development and broader dissemination of techniques at the different stages of a molecular data mining workflow.
      
    \end{itemize}

\bibliography{44_ML_review}
\end{document}